\documentclass{article}
\usepackage{color}
\usepackage{algorithm}
\usepackage{algpseudocode}
\usepackage{amsmath,amssymb,amsfonts}
\usepackage{booktabs}       
\usepackage{multirow, varwidth}
\usepackage{mathtools}
\usepackage{epsfig}
\usepackage{verbatim}
\usepackage{xr}
\usepackage{flushend}
\usepackage{subcaption}

\usepackage{graphicx}
\usepackage{url}
\usepackage{enumitem}
\setlist{nosep} 
\usepackage[table]{xcolor}         
\usepackage{kotex}
\usepackage{colortbl}

\DeclareRobustCommand\onedot{\futurelet\@let@token\@onedot}
\def\onedot{.} 
\def\eg{\emph{e.g}\onedot, }

 \def\vs{\emph{vs}\onedot}

\newcommand{\mytilde}{\raise.17ex\hbox{$\scriptstyle\mathtt{\sim}$}}

\usepackage{spconf,amsmath,graphicx,hyperref}
\usepackage{amssymb, pifont}

\title{SIREN: Spatially-Informed Reconstruction of Binaural Audio with Vision}
\name{Mingyeong Song$^*$, Seoyeon Ko$^*$, Junhyug Noh
\thanks{$^*$Equal contribution}
}
\address{
Ewha Womans University, Seoul, Korea
}
\begin{document}
\ninept
\maketitle
\begin{abstract}
Binaural audio delivers spatial cues essential for immersion, yet most consumer videos are monaural due to capture constraints. We introduce \textbf{SIREN}, a visually guided mono$\rightarrow$binaural framework that \emph{explicitly} predicts left and right channels. A ViT-based encoder learns dual-head self-attention to produce a shared scene map and end-to-end L/R attention -- replacing hand-crafted masks. A soft, annealed spatial prior gently biases early L/R grounding, and a two-stage, confidence-weighted \emph{waveform}-domain fusion (guided by mono reconstruction and interaural phase consistency) suppresses crosstalk when aggregating multi-crop and overlapping windows. Evaluated on FAIR-Play and MUSIC-Stereo, SIREN yields consistent gains on time–frequency and phase-sensitive metrics with competitive SNR. The design is modular and generic, requires no task-specific annotations, and integrates with standard audio-visual pipelines.
\end{abstract}

\section{Introduction}
\label{sec:intro}

Humans localize sound by exploiting interaural time and level differences (ITD/ILD), which enables stable spatial perception and immersion in VR/AR, gaming, and interactive media~\cite{hoeg2017binaural}. 
However, most consumer videos provide only mono audio, since capturing true binaural signals typically requires specialized microphones and careful recording setups.

To bridge this gap, recent work uses synchronized video as spatial context to convert mono audio into binaural audio~\cite{Gao201825DVS,Xu2021VisuallyIB,Rachavarapu2021LocalizeTB,Zhou2020SepStereoVG,Parida2021BeyondMT,Li2024CyclicLF,Liu2024VisuallyGB,Chen2025CCStereoAC}. 
Despite steady progress, three challenges remain: (i) conventional visual backbones may miss fine-grained spatial evidence needed for reliable left/right (L/R) formation; (ii) L/R guidance is often imposed manually or via rigid heuristics rather than learned end-to-end; and (iii) test-time aggregation over multi-crop and overlapping windows can introduce L/R leakage and unstable spatial cues.

We introduce \textbf{SIREN} (Spatially-Informed REconstruction of binaural audio with visioN), a visually guided mono$\rightarrow$binaural framework that \emph{explicitly} predicts L/R channels while learning directional cues within a transformer. 
Specifically, a Vision Transformer (ViT) encoder learns dual-head self-attention over patch tokens to produce a shared scene map and L/R attention maps end-to-end, replacing hand-crafted L/R masks. 
To stabilize early grounding, we add a soft spatial prior that is annealed away over training. 
At inference, we further apply a two-stage confidence-weighted waveform fusion guided by mono reconstruction and interaural-phase consistency, which reduces L/R leakage when combining multi-crop and overlapping predictions.

Our main contributions are as follows:
\begin{itemize}[leftmargin=*]
\item \textbf{Explicit L/R prediction via transformer-native attention:} using a ViT-based encoder, we learn dual-head self-attention that yields a shared scene map and L/R-specific attention maps end-to-end, enabling channel-wise refinement without hand-crafted heuristics.
\item \textbf{Soft spatial prior for robust L/R grounding:} logistic target maps used as decaying supervision provide an initial directional bias and smoothly vanish to allow content-driven learning.
\item \textbf{Confidence-weighted test-time refinement:} a two-stage waveform-domain fusion guided by mono and interaural-phase consistency mitigates L/R leakage and stabilizes aggregation across crops and overlapping windows.
\item \textbf{Comprehensive evaluation:} experiments on FAIR-Play and MUSIC-Stereo demonstrate consistent gains on standard fidelity and spatial metrics.
\end{itemize}

\section{Related Work}
\label{sec:related_work}

\noindent\textbf{Audio-visual binaural audio generation.}
Visually guided binaural spatialization was introduced by~\cite{Gao201825DVS} and has evolved from addressing data scarcity~\cite{Xu2021VisuallyIB} to improving spatial fidelity via visual pre-localization of sound sources~\cite{Rachavarapu2021LocalizeTB,Zhou2020SepStereoVG}. 
Localize-then-synthesize pipelines are particularly effective in multi-source scenes~\cite{Zhou2020SepStereoVG}, while other directions strengthen audio--visual coupling using depth cues~\cite{Parida2021BeyondMT}, cycle consistency~\cite{Li2024CyclicLF}, or multi-task objectives~\cite{Li2021BinauralAG}. 
To support broader generalization, recent work has also introduced larger-scale datasets spanning diverse indoor/outdoor conditions~\cite{Li2024BinauralmusicAD}. 
Our method is most related to CMC~\cite{Liu2024VisuallyGB}, which predicts a difference spectrogram via modality alignment; in contrast, we explicitly predict left/right channels with transformer-native L/R attention and stabilize inference with confidence-weighted fusion. 
Concurrently, contrastive learning and refined inference have further improved mono-to-binaural generation~\cite{Chen2025CCStereoAC}.

\smallskip
\noindent\textbf{Self-supervised visual representations.}
Self-supervised learning (SSL) has become a strong source of visual features for audio--visual tasks.
Backbones such as DINO~\cite{Caron2021EmergingPI} and MAE~\cite{He2021MaskedAA} benefit problems including sounding object segmentation~\cite{Bhosale2024UnsupervisedAS,Caron2021EmergingPI} and sound localization~\cite{Radford2021LearningTV}. 
Audio--visual SSL models (\eg AudioCLIP~\cite{Guzhov2021AudioclipEC} and AV-MAE~\cite{Georgescu2022AudiovisualMA}) further support this trend. 
Motivated by these advances, we adopt DINOv3~\cite{Simeoni2025DINOv3} as a ViT-based encoder to provide fine-grained visual evidence for visually guided binaural generation.

\begin{figure*}[t!]
    \centering 
    \includegraphics[width=0.99\textwidth]{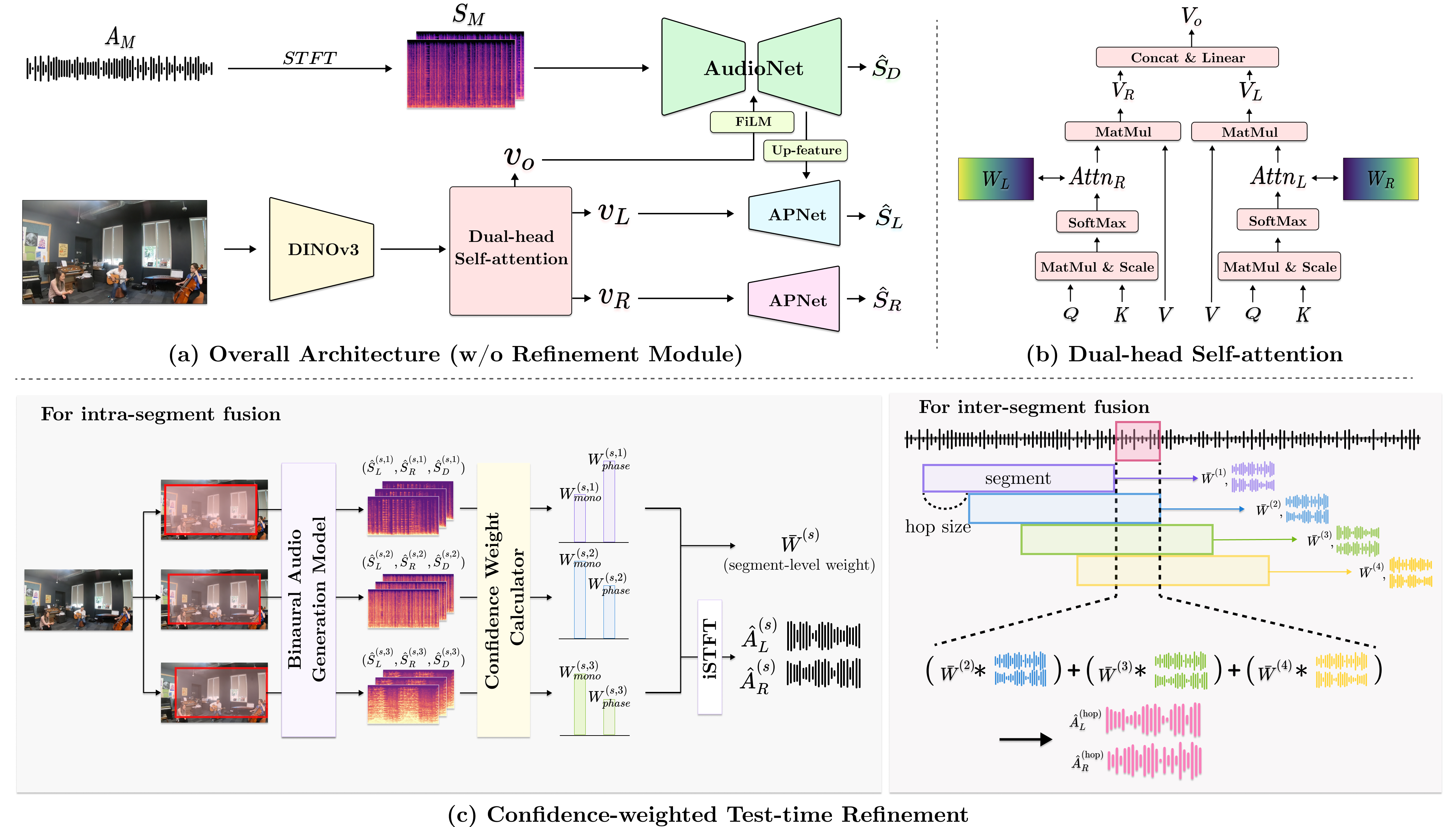}
    \caption{
    \textbf{Overview.} (a, b) SIREN takes mono STFT $S_M$ and video $V$; a ViT with dual-head attention produces a shared map $v$ and L/R features that FiLM-condition an audio U-Net to predict $\hat{S}_D$ and $(\hat{S}_L,\hat{S}_R)$. (c) At test time, two-stage confidence-weighted refinement fuses crop-wise and overlap-wise waveforms (post-iSTFT) using mono and interaural-phase scores.
    }
    \label{fig:architecture} 
\end{figure*}

\section{Method}
\label{sec:method}

Given mono audio $A_M$ synchronized with video $V$, our goal is to reconstruct binaural signals $(\hat{S}_L,\hat{S}_R)$. At a high level (Fig.~\ref{fig:architecture}), SIREN couples transformer-native L/R attention with FiLM-conditioned audio generation and channel-wise refinement; a soft, annealed spatial prior gently biases early L/R grounding, and a lightweight confidence-weighted waveform fusion stabilizes test-time aggregation. 

\subsection{Preliminaries}
\label{sec:prelim}
Let $A_L, A_R \in \mathbb{R}^{T}$ be left/right waveforms and $A_M=A_L+A_R$ their mono mix. 
Applying STFT yields complex spectrograms $S_L, S_R, S_M \in \mathbb{C}^{F\times U}$ over frequency bins $F$ and time frames $U$. 
We additionally define the difference spectrogram
\begin{equation}
S_D = S_L - S_R ,
\label{eq:diffspec}
\end{equation}
and during training predict an auxiliary $\hat{S}_D$ to enhance spatial reconstruction.

Formally, given $S_M$ and video frames $V$,
\begin{equation}
(\hat{S}_L,\,\hat{S}_R) = F(S_M, V),
\label{eq:mapping}
\end{equation}
and the corresponding waveforms are recovered via iSTFT.

\subsection{Model Architecture}
\label{sec:model_arc}

\noindent\textbf{Vision feature encoder with dual-head self-attention.}
A ViT-based encoder (DINOv3~\cite{Simeoni2025DINOv3}) operates on patch tokens and, \emph{within the transformer}, learns \emph{dual-head self-attention} that naturally yields left/right selectivity -- replacing hand-crafted L/R masks with end-to-end, data-driven directional cues. 
Tokens are projected back to a spatial feature map $v\!\in\!\mathbb{R}^{C\times H\times W}$, while the two heads produce softmax-normalized attention maps $\mathrm{Attn}_L,\mathrm{Attn}_R\!\in\![0,1]^{H\times W}$. L/R-specific features are formed by modulating the shared map $v$:
\begin{equation}
v_L = v \odot \mathrm{Attn}_L,\qquad v_R = v \odot \mathrm{Attn}_R,
\end{equation}
with channel-wise broadcasting over $v$. Thus $v$ captures scene semantics for shared conditioning, and $(v_L,v_R)$ provide transformer-native directional guidance to the audio heads. A global descriptor for FiLM is obtained by pooling $v$ (\eg global average) and passing it through a small MLP.

\smallskip
\noindent\textbf{Audio U-Net with FiLM conditioning.}
The U-Net encoder ingests the \emph{complex} mono spectrogram by stacking its real and imaginary parts as two channels,
\begin{equation}
X_M=\big[\Re(S_M)\,;\,\Im(S_M)\big],
\end{equation}
so $(\Re(S_M),\Im(S_M))$ denote the real/imaginary (quadrature) components of the STFT, preserving magnitude \emph{and} phase information. The decoder reconstructs multi-resolution audio features; at each upsampling stage, FiLM layers~\cite{Perez2017FiLMVR} modulate feature statistics using the pooled descriptor from $v$. The decoder outputs (i) an initial estimate $\hat{S}_D$ and (ii) a multi-resolution pyramid forwarded to the channel heads.

\smallskip
\noindent\textbf{Channel refinement heads.}
Two refinement heads (APNet-style~\cite{Zhou2020SepStereoVG}) take the pyramid together with $v_L$ (left head) or $v_R$ (right head) and emit complex spectrograms $\hat{S}_L$ and $\hat{S}_R$. Separating the shared conditioning (from $v$) from directional cues ($v_L,v_R$) improves spatial specificity.

\subsection{Spatial Prior Loss}
\label{sec:prior}
We encourage opposing left/right biases \emph{early} in training by aligning the learned attention maps to simple directional targets; the weight of this prior decays to zero as training proceeds. The target maps follow the \emph{logistic ramp} design popularized in CMC~\cite{Liu2024VisuallyGB} (there used directly as attention), while we employ them as \emph{decaying supervision signals}.
Let $x\!\in\!\{1,\dots,W\}$ denote the column index. Define
\begin{align*}
W_{L}(:, :, x) = \frac{1}{1+e^{qx-r}},\quad
W_{R}(:, :, x) = \frac{1}{1+e^{-qx-r}},
\end{align*}
and normalize so $\sum_{h,w}W_{\{L,R\}}(h,w)=1$. With learned, normalized maps $\mathrm{Attn}_L,\mathrm{Attn}_R$, the prior is
\begin{equation*}
\mathcal{L}_{\text{prior}}(t)
= \lambda_{\text{prior}}(t)\Big[
\operatorname{MSE}(\mathrm{Attn}_L,W_L)+\operatorname{MSE}(\mathrm{Attn}_R,W_R)\Big],
\end{equation*}
where $\lambda_{\text{prior}}(t) = \lambda_0\,\max(0,\,1-t/T_{\text{anneal}})$.
This guidance provides an initial ``push'' toward directional grounding while allowing content-driven spatial learning later.

\subsection{Training Objectives}
\label{sec:losses}
We add two reconstruction losses and form the total objective as a weighted sum:
\begin{align}
\mathcal{L}_{D} &= \big\| \hat{S}_D - S_D \big\|_2, \\
\mathcal{L}_{RL} &= \big\| \hat{S}_R - S_R \big\|_2 + \big\| \hat{S}_L - S_L \big\|_2, \\
\mathcal{L} &= \mathcal{L}_{D} + \lambda_{RL}\mathcal{L}_{RL} + \lambda_{\text{prior}}\mathcal{L}_{\text{prior}}.
\end{align}

\subsection{Confidence-Weighted Test-time Refinement}
\label{sec:refine}

Inference runs on fixed-length \emph{segments} (\eg 0.63\,s) with a sliding window and hop $H$, so adjacent segments overlap.
Let a \emph{hop frame} denote the minimal overlap unit aligned to the hop (one hop long).
Each hop frame is covered by $N$ different segments; the naive approach averages their predictions equally, ignoring quality differences and often causing timbral drift or spatial collapse.
We therefore adopt a two-stage refinement using universal, physically motivated confidence scores.
For any binaural candidate $(\hat{S}_L,\hat{S}_R,\hat{S}_D)$ we compute two complementary scores:

\smallskip
\noindent\textbf{Mono consistency.}
Physically, the mono input is a linear mixture of the ears. Thus a binaural hypothesis whose \emph{predicted mono} matches the input $S_M$ is less likely to introduce spectral tilt or band-wise holes.
We form the predicted mono (a scale factor like $1/2$ only rescales errors and is absorbed by normalization): $\hat{S}_{M}=(\hat{S}_L+\hat{S}_R)/2$.
We then measure a magnitude-domain discrepancy and convert it to a weight:
\begin{equation}
e_{\text{mono}}=\operatorname{mean}\!\big(\,\big||\hat{S}_M|-|S_M|\big|\,\big),\quad
W_{\text{mono}}=\frac{1}{e_{\text{mono}}+\epsilon}.
\end{equation}

\smallskip
\noindent\textbf{Interaural phase consistency.}
Interaural phase/ITD cues are important for lateralization, and we use phase agreement as a \emph{signal-level} reliability check during test-time selection.
Specifically, we compare the phase of the \emph{reconstructed} difference signal to the auxiliary difference branch.
Let $\tilde{S}_D=\hat{S}_L-\hat{S}_R$ and $\arg(S)=\operatorname{atan2}(\Im(S),\Re(S))$. We compute
\begin{equation}
e_{\text{phase}}=\big|\arg(\tilde{S}_D)-\arg(\hat{S}_D)\big|,\quad
W_{\text{phase}}=\frac{1}{e_{\text{phase}}+\epsilon}.
\end{equation}
Although perceptual IPD is most relevant below $\sim$1.5\,kHz, we evaluate $e_{\text{phase}}$ over all frequencies because it measures \emph{self-consistency} between two difference-signal estimates, helping reject candidates with broadband phase artifacts.

\smallskip
\noindent\textbf{Product-of-experts weight.}
Neither criterion alone suffices; they trade off. We therefore combine them multiplicatively:
\begin{equation}
W=W_{\text{mono}}\times W_{\text{phase}},
\end{equation}
favoring candidates that are \emph{simultaneously} timbrally faithful and spatially consistent.
Based on this unified score $W$, we perform two stages: (i) intra-segment fusion over $K$ visual crops to obtain a single time-domain (iSTFT) waveform per segment, and (ii) inter-segment fusion across the $N$ segment-level waveforms that cover each hop frame. We detail both stages below.

\smallskip
\noindent\textbf{Stage 1: Intra-segment fusion over $K$ crops.}
For each segment $s$, generate $K{=}3$ crop-wise candidates $\{(\hat{S}_L^{(s,k)},\hat{S}_R^{(s,k)},\hat{S}_D^{(s,k)})\}_{k=1}^{K}$ and compute $W^{(s,k)}$ as above. Normalize within the segment,
\begin{equation}
\bar{W}^{(s,k)}=\frac{W^{(s,k)}}{\sum_{k'=1}^{K} W^{(s,k')}}.
\end{equation}
Convert each candidate to time-domain waveforms via iSTFT,
\begin{equation}
\hat{A}_L^{(s,k)}=\operatorname{iSTFT}\!\big(\hat{S}_L^{(s,k)}\big),\quad
\hat{A}_R^{(s,k)}=\operatorname{iSTFT}\!\big(\hat{S}_R^{(s,k)}\big),
\end{equation}
and obtain a single \emph{segment-level waveform} by weighted summation:
\begin{equation}
\hat{A}_L^{(s)}=\sum_{k=1}^{K}\bar{W}^{(s,k)}\,\hat{A}_L^{(s,k)},\quad
\hat{A}_R^{(s)}=\sum_{k=1}^{K}\bar{W}^{(s,k)}\,\hat{A}_R^{(s,k)}.
\end{equation}

\smallskip
\noindent\textbf{Stage 2: Inter-segment fusion per hop frame.}
Let a given hop frame be covered by segments $\mathcal{N}$ with $|\mathcal{N}|=N$. 
Treat each segment-level prediction $(\hat{S}_L^{(s)},\hat{S}_R^{(s)},\hat{S}_D^{(s)})$, $s\in\mathcal{N}$, as a candidate and \emph{reuse the same scoring}. Normalize across segments,
\begin{equation}
\bar{W}^{(s)}=\frac{W^{(s)}}{\sum_{s'\in\mathcal{N}} W^{(s')}}.
\end{equation}
Fuse \emph{waveforms} over the hop-frame interval (restriction denoted by $|_{\text{hop}}$), then perform standard overlap–add:
\begin{equation}
\hat{A}_L^{(\text{hop})}=\sum_{s\in\mathcal{N}}\bar{W}^{(s)}\,\hat{A}_L^{(s)}\Big|_{\text{hop}},\,\,\,
\hat{A}_R^{(\text{hop})}=\sum_{s\in\mathcal{N}}\bar{W}^{(s)}\,\hat{A}_R^{(s)}\Big|_{\text{hop}}.
\end{equation}
Assembling all hop frames yields the final binaural waveform.

\begin{table*}[t]
\centering
\footnotesize 
\small
\setlength{\tabcolsep}{10pt}
\caption{Comparison with prior work on FAIR-Play (10-split) and MUSIC-Stereo. 
Lower is better for STFT/ENV/Phs; higher is better for SNR. ``--'' denotes not reported.}
\label{tab:results}
\begin{tabular}{l|cccc|cccc}
\toprule
\multirow{2}{*}{Method} 
& \multicolumn{4}{c|}{FAIR-Play (10-split)} 
& \multicolumn{4}{c}{MUSIC-Stereo} \\
\cmidrule(lr){2-5}\cmidrule(lr){6-9}
& STFT $\downarrow$ & ENV $\downarrow$ & Phs $\downarrow$ & SNR $\uparrow$
& STFT $\downarrow$ & ENV $\downarrow$ & Phs $\downarrow$ & SNR $\uparrow$ \\
\midrule
Mono2Binaural~\cite{Gao201825DVS}& 0.889 & 0.137 & 1.438 & 6.232 & 0.942 & 0.138 & 1.550 & 8.255 \\
Sep-Stereo~\cite{Zhou2020SepStereoVG} 
& 0.879 & 0.136 & 1.429 & -- & 0.929 & 0.135 & 1.544 & 8.306 \\
CMC~\cite{Liu2024VisuallyGB} & 0.849 & 0.133 & \textbf{1.423} & -- & 0.759 & 0.113 & 1.502 & -- \\
CC-Stereo~\cite{Chen2025CCStereoAC} 
& 0.823 & 0.132 & -- & 7.144 & 0.624 & 0.097 & 1.578 & \textbf{12.985} \\
\textbf{SIREN (Ours)} & \textbf{0.820} & \textbf{0.132} & 1.550 & \textbf{7.219}
& \textbf{0.417} & \textbf{0.091} & \textbf{1.006} & 10.872 \\
\bottomrule
\end{tabular}
\end{table*}

\section{Experiments}
\label{sec:exp}

\begin{table}[t]
\centering
\small
\setlength{\tabcolsep}{8.2pt}
\caption{Ablation on FAIR-Play (5-split): impact of the spatial prior loss ($L_{\text{prior}}$) and confidence-weighted refinement ($W$).}
\label{tab:ablation}
\begin{tabular}{cc|cccc}
\toprule
$L_{\text{prior}}$ & $W$ & STFT $\downarrow$ & ENV $\downarrow$ & Phs $\downarrow$ & SNR $\uparrow$ \\
\midrule
\ding{55} & \ding{55} & 0.941 & 0.141 & 1.599 & 6.345 \\
\ding{51} & \ding{55} & 0.928 & 0.140 & \textbf{1.584} & 6.224 \\
\ding{51} & \ding{51} & \textbf{0.888} & \textbf{0.136} & 1.589 & \textbf{6.798} \\
\bottomrule
\end{tabular}
\end{table}

\subsection{Experimental Settings}
\label{sec:exp_setting}

\noindent\textbf{Datasets.}
We evaluate on two audio--visual benchmarks widely used in prior work:
\begin{itemize}[leftmargin=*]
\item \textbf{FAIR-Play~\cite{Gao201825DVS}.} 1,871 clips of 10\,s ($\sim$5.2\,h) recorded with a professional binaural microphone setup. We report results on the official 10-split (train/val/test = 1,497/187/187) and a more challenging 5-split protocol~\cite{Xu2021VisuallyIB}. Video is sampled at 10\,fps.
\item \textbf{MUSIC-Stereo~\cite{zhao2018sound, zhao2019sound}.} Built from MUSIC (21 instrument classes). We use 720 videos filtered for clear interaural difference (avg. L--R difference $>0.001$)~\cite{Chen2025CCStereoAC}, yielding 15{,}302 clips of 10\,s each, with an 80/10/10 train/val/test split and 10\,fps video.\footnote{A small subset of source videos may be unavailable over time, so the exact clip composition can differ slightly from prior work; we follow the same filtering and splitting protocol.}
\end{itemize}

\noindent\textbf{Implementation details.}
For each 10\,s clip, we randomly crop a 0.63\,s audio segment (10{,}080 samples at 16\,kHz; $\approx$63 STFT frames) and pair it with the central video frame. 
Audio is resampled to 16\,kHz and transformed via a \emph{complex} STFT (Hann; window=512, hop=160\,(10\,ms), FFT=512); the network ingests $[\Re(S_M);\Im(S_M)]$. Video frames are resized to $480{\times}240$ and randomly cropped to $448{\times}224$. 
We train a U\!-Net encoder–decoder for audio and use DINOv3 ViT-B/16 for visual features, with batch size 32 and loss weights $\lambda_{\text{prior}}{=}2$ and $\lambda_{RL}{=}5$. 
At inference, we use a 0.63\,s sliding segment with hop $H{=}0.05$\,s ($\approx$13 overlaps) and apply the confidence-weighted refinement described in Sec.~\ref{sec:refine}; the final binaural waveform is obtained by overlap–add and iSTFT.


\noindent\textbf{Evaluation Metrics.}
We adopt four standard metrics widely used in prior works to assess signal fidelity and spatial quality:
\begin{itemize}[leftmargin=*]
    \item \textbf{STFT L2 (STFT):} $\ell_2$ distance between complex spectrograms, measuring time--frequency consistency.
    \item \textbf{Envelope Distance (ENV):} $\ell_2$ distance between time-domain envelopes, reflecting perceptual similarity.
    \item \textbf{Phase Distance (Phs):} $\ell_1$ distance between STFT phases, evaluating phase accuracy in the time--frequency domain.
    \item \textbf{Signal-to-Noise Ratio (SNR):} Ratio of ground truth to error signal power (dB), assessing overall signal quality.
\end{itemize}

\subsection{Results}
\label{sec:results}

\noindent\textbf{Main results.}
Table~\ref{tab:results} summarizes performance on FAIR-Play (10-split) and MUSIC-Stereo. Overall, SIREN improves time--frequency fidelity and spatial stability: on MUSIC-Stereo it achieves the lowest STFT/ENV/Phs, indicating consistently better reconstruction across spectral magnitude, temporal envelope, and phase. On FAIR-Play, SIREN attains the best STFT and SNR, with a modest Phs gap to CMC ($1.550$ \vs\ $1.423$). We attribute this gap to explicit L/R prediction combined with waveform-domain fusion and a mono-consistency bias, which can leave minor interaural phase mismatches in controlled recordings; increasing the phase weight or adding IPD/group-delay terms may mitigate it.

\smallskip
\noindent\textbf{Prior and refinement.}
Table~\ref{tab:ablation} isolates the contributions of the spatial prior loss and the confidence-weighted refinement. The prior mainly improves Phs by encouraging an early left/right bias, which stabilizes directional grounding before the model fully relies on content-driven cues. In contrast, the refinement consistently improves STFT/ENV/SNR by selecting and reweighting crop-wise and overlap-wise predictions, reducing leakage and aggregation artifacts. Enabling both yields the best overall trade-off, combining stronger phase behavior with improved reconstruction fidelity.

\smallskip
\noindent\textbf{Dual-head attention and FiLM.}
Table~\ref{tab:ablation2} examines the roles of dual-head attention and FiLM conditioning (with the refinement fixed). Removing dual-head attention forces both ears to share the same visual evidence, weakening channel-specific guidance and degrading STFT/ENV. Removing FiLM eliminates multi-scale visual modulation inside the audio decoder, which reduces the benefit of global scene conditioning and hurts overall quality. Together, these results support that dual-head attention provides ear-specific spatial cues while FiLM supplies complementary global conditioning, and that both are important for robust binaural reconstruction.



\begin{table}[t]
\centering
\small
\setlength{\tabcolsep}{6.5pt}
\caption{Ablation on FAIR-Play (5-split): effect of FiLM conditioning and dual-head attention.}
\label{tab:ablation2}
\begin{tabular}{cc|cccc}
\toprule
FiLM & DualHead & STFT $\downarrow$ & ENV $\downarrow$ & Phs $\downarrow$ & SNR $\uparrow$ \\
\midrule
\ding{55} & \ding{55} & 0.935 & 0.141 & 1.582 & 6.379 \\
\ding{51} & \ding{55} & 0.925 & 0.140 & \textbf{1.576} & 6.432 \\
\ding{55} & \ding{51} & 0.913 & 0.139 & 1.577 & 6.475 \\
\ding{51} & \ding{51} & \textbf{0.888} & \textbf{0.136} & 1.589 & \textbf{6.798} \\
\bottomrule
\end{tabular}
\end{table}

\section{Conclusion}
\label{sec:conclusion}
We presented SIREN, a visually guided mono-to-binaural framework that \emph{explicitly} predicts left and right channels, moving beyond difference--spectrogram formulations. 
By combining a DINOv3 ViT encoder with transformer-native dual-head L/R attention, SIREN provides end-to-end directional visual cues for audio generation, while a soft spatial prior and confidence-weighted test-time refinement improve spatial stability and reduce crosstalk during aggregation. 
Experiments on FAIR-Play and MUSIC-Stereo show strong performance across standard fidelity and spatial metrics, demonstrating that explicit L/R prediction with principled inference yields high-quality, spatially coherent binaural audio.

\smallskip
\noindent\textbf{Acknowledgements.}
This work was supported by the National Research Foundation of Korea (NRF) grant funded by the Korea government (MSIT) (No. RS-2025-16070597) and Global - Learning \& Academic research institution for Master’s·PhD students, and Postdocs (G-LAMP) Program of the National Research Foundation of Korea (NRF) grant funded by the Ministry of Education (No. RS-2025-25442252).

\bibliographystyle{IEEEbib}
\bibliography{refs}

\end{document}